\documentstyle[preprint,revtex]{aps}

\def\ap#1#2#3{           {\it Ann. Phys. (NY) }{\bf #1}, #2 (19#3)}

\def\fp#1#2#3{           {\it Fort. der Physik }{\bf #1}, #2 (19#3)}

\def\np#1#2#3{           {\it Nucl. Phys. }{\bf #1}, #2 (19#3)}
\def\pl#1#2#3{           {\it Phys. Lett. }{\bf #1}, #2 (19#3)}
\def\pr#1#2#3{           {\it Phys. Rev. }{\bf #1}, #2 (19#3)}

\newcommand{\bea}{\begin{eqnarray}}
\newcommand{\eea}{\end{eqnarray}}
\newcommand{\beq}{\begin{equation}}
\newcommand{\eeq}{\end{equation}}

\begin{document}
\preprint{IFUSP/P-1001}
\preprint{ hep-ph/9208234 }
\draft

\begin{title}
On the Infrared Behavior of the Pressure\\
in Thermal Field Theories
\end{title}

\author{A. P. de Almeida and J. Frenkel}
\begin{instit}
Instituto de F\'\i sica, Universidade de S\~ao Paulo,\\
S\~ao Paulo, 01498 SP, Brasil
\end{instit}

\begin{abstract}
We study non-perturbatively, via the Schwinger-Dyson equations, the leading
infrared behavior of the pressure in the ladder approximation. This problem is
discussed firstly in the context of a thermal scalar field theory, and the
analysis is then extended to the Yang-Mills theory at high temperatures. Using
the Feynman gauge, we find a system of two coupled integral equations for the
gluon and ghost self-energies, which is solved analytically. The solutions of
these equations show that the contributions to the pressure, when calculated in
the ladder approximation, are finite in the infrared domain.
\end{abstract}

\section{Introduction}

Relativistic field theories at finite temperature have been actively studied in
the past years, because of their relevance to the theory of the early universe
and to the quark-gluon plasma which may be created in heavy ion
collisions\cite{ref1}. The study of such physical problems requires the
computation of the thermodynamic potential, from which other thermodynamic
properties, like the pressure, may be determined. It was pointed out by
Linde\cite{ref2} that the thermodynamic potential of the Yang-Mills theory
cannot be calculated beyond the fifth order of the coupling constant. This is
related with the infrared singularities of the non-abelian gauge theories at
finite temperatures, which arise from
the fact that the magnetic mass vanishes at
least up to second order in perturbation theory. The infrared problem in
these theories at finite temperature is qualitatively different from that at
$T=0$, manifesting itself by the presence of infrared power divergences in
higher orders of the perturbative expansion.

One may hope that the use of non-perturbative methods can throw some light on
this problem. Such methods generally start with a discussion of the relevant
Schwinger-Dyson equations\cite{ref3}. It is well known that this set of
integral equations is not closed since an n-point function is generally related
to the $n+1$ and $n+2$ functions. This is what makes the solutions of
such problems exceedingly difficult. For this reason one must resort in
practice to some approximations, which result from truncating in some way the
set of the Schwinger-Dyson equations. This non-perturbative approach has been
used by Jackiw and Templeton\cite{ref4}, to show how super-renormalizable
interactions might cure their infrared divergences. It has also been employed
by Mandelstam\cite{ref5} to analyze the infrared behavior of the gluon
propagator,using the ladder approximation at $T=0$.
He showed that the ensuing set of the Schwinger-Dyson
equations does provide confinement in the strong coupling regime.

The purpose of this work is to study, in the ladder approximation, the leading
infrared behavior of the pressure in the Yang-Mills theory
at high temperatures. Our
approach, via the Schwinger-Dyson equations,
is similar in spirit with that advocated by Kajantie and
Kapusta\cite{ref6}. Namely, we regard the simplified set of these
equations just as a convenient procedure for summing particular classes of an
infinite number of diagrams in the weak coupling regime. In order to regularize
the theory in the event that no magnetic masses are generated, we shall put an
infrared cut-off $\lambda$ on the momenta integrations which arise when
calculating the contributions to the pressure. Our aim is to study the
possibility of cancellation of the infrared singularities in the limit
$\lambda\rightarrow0$, when summing the whole class of diagrams which
contribute in the ladder approximation.

As we shall see, even
this more modest task is non-trivial and we begin by considering in Section
\ref{scalar} the
scalar $g \phi^3$ theory in six dimensions. This model has some
similarities with the Yang-Mills theory, such as a dimensionless coupling
constant and asymptotic freedom. Strictly speaking, in this case one could
employ the re-summation method developed by Braaten and Pisarski\cite{ref8},
because a scalar thermal mass is generated in lowest order. Indeed, such a
procedure leads to an infrared finite expression for the pressure\cite{ref9}.
However, we will use here instead the method described above, in order to
illustrate in a simple way some relevant features which will appear in
the Yang-Mills
theory. We consider the truncated set of Schwinger-Dyson equations which yield
consistently the scalar self-energy in the ladder approximation. The ensuing
integral equation is equivalent, under certain conditions to be discussed
later,
to a second order differential equation which can be solved analytically
in
terms of modified Bessel functions. Using this solution, we show that the
corresponding contribution to the pressure obtained in this approximation, is
finite in the limit $\lambda \rightarrow 0$.

In Section \ref{ym} we consider the leading infrared behavior of the
$SU(N)$ Yang-Mills theory at
high temperatures , in the ladder approximation. We work in the Feynman gauge
and obtain a system of two coupled Schwinger-Dyson integral equations, linking
the gluon polarization tensor and the ghost self-energy function. Using the
same reasoning as in the previous section, we show that this system is
equivalent to a fourth order differential equation which describes in this
approximation the leading infrared behavior. The solution of this equation is
expressed in terms of modified Bessel functions with complex argument and the
corresponding complex conjugate functions. We find that the
relevant
contributions to the pressure, as calculated in the ladder approximation, are
finite in the infrared limit $\lambda \rightarrow 0$. Some of the mathematical
details which arise during these calculations are given in the Appendices
\ref{aa} and \ref{bb}. Finally, in Appendix \ref{cc} we consider the
possibility of the cancellation of the infrared divergences in the pressure,
beyond the ladder approximation.

\section{The Six-Dimensional Scalar Theory}
\label{scalar}

We consider here the scalar theory with a $g \phi^3$ interaction at the
temperature $T$. In this theory the formula giving the pressure in the ladder
approximation is rather simple and can be expressed directly in terms of the
self-energy function $\tilde\Pi$. The Schwinger-Dyson equation for the scalar
self-energy in the ladder approximation is shown in Fig. (1). Note that the
diagrams contributing to $\tilde\Pi$ have the same combinatorial factor
$\frac12$ to all orders in perturbation theory. The corresponding contributions
to the pressure are represented in Fig. (2.a).

In order to derive the expression giving the pressure, we use the relation
between its functional derivative and the self-energy function\cite{ref1}:

\beq
\label{f1}
\left(\frac{\delta P}{\delta{\cal D}_0}\right)_{\rm 1PI}=-\frac T2\tilde\Pi
\eeq
where ${\cal D}_0$ is the free particle propagator and ${\rm 1PI}$ indicates
that only one particles irreducible diagrams contribute to $\tilde\Pi$. With
the help of the graphical representation given in Fig. (2.a), it is easy to see
that (\ref{f1}) implies the relation:

\beq
\label{f2}
P=P_0-\frac T2\frac14\sum_{p_0}\int\frac{d^5p}{(2\pi)^5}\frac1{p^2+p_0^2}
\left[\tilde\Pi(p,p_0)+\frac13\tilde\Pi^{(1)}(p,p_0)\right]
\eeq

The factor $\frac14$ arises because in general there are only four ways we can
cut the diagram (2.a) in order to get $\tilde\Pi$ in the ladder approximation.
The only exception occurs in the lowest order, where the factor is
$\frac13$ instead. The graphical representation of this formula is shown in
Fig. (2.b).

The sum over $p_0$ should be taken over even frequencies $p_0=2\pi nT$.
However, the dominant infrared contributions arise only from the terms with
zero frequency ($n=0$). When calculating the contributions from eq. (\ref{f2})
corresponding to this mode, we can put an ultraviolet cut-off of order $T$ on
the momentum integration. This cut-off arise naturally when summing over all
modes $n$. As mentioned in the Introduction, we shall also put a cut-off
$\lambda$ in order to regularize the perturbative infrared divergences. In this
way we find that the leading infrared contributions to the pressure are given
by:

\beq
\label{tl}
P=P_0-\frac T8\int_\lambda^T\frac{d^5p}{(2\pi)^5}\frac1{p^2}\tilde\Pi(p,p_0=0)
+\dots
\eeq
where dots denote additional subleading and infrared convergent
contributions.

Then, using the Schwinger-Dyson equation for the scalar self-energy
function we arrive at the following integral equation for
$\tilde\Pi(p,p_0=0)$:

\beq
\label{f4}
\tilde{\Pi}(p)\equiv\Pi(p)+p^2=-\frac{g^2T}{(2\pi)^5}
\int_{\lambda}^T\frac{d^5 k}{({\bf p}+{\bf k})^2}
\left[\frac1{2k^2}+\frac{\Pi(k)}{k^4}
\right]
\eeq

In order to solve this integral equation analytically, we will try to convert
it into a Volterra type. This is possible provided we adopt the following
maximization procedure. Let us define:

\beq
({\bf p}+{\bf k})^2_{\rm max}\equiv\left\{
\begin{array}{ll}
p^2\quad&{\rm for}\ p>k\\
k^2\quad&{\rm for}\ k>p
\end{array}
\right.
\eeq

Using this approach, we then obtain from eq. (\ref{f4}):

\beq
\label{f7}
\Pi_{\rm max}(p)=-p^2
+\frac{g^2T}{12\pi^3}\left[\frac p3-\frac T2+\frac16\frac{\lambda^3}{p^2}
-\frac{1}{p^2}\int_\lambda^pdk\Pi_{\rm max}(k)
-\int_p^T \frac{dk}{k^2}\Pi_{\rm max}(k)\right]
\eeq

Comparing this with the original equation (\ref{f4}), we note that the
first iteration of (\ref{f7}) with $\lambda\rightarrow0$ yields:

\beq
\Pi^{(1)}_{\rm max}=-p^2+\frac{g^2 T^2}{24
\pi^3}\left[1-\frac{2}{3}\frac{p}{T}\right]
\eeq
a result which completely agrees in the high temperature limit with the
leading contribution which follows from (\ref{f4}). To next order, our
procedure
yields in this regime results which are in satisfactory agreement with the ones
obtained from equation (\ref{f4}), which becomes increasingly cumbersome to
handle . Since we wish to study only the main features of the leading
infrared behavior, it will be sufficient for our purpose to restrict our
attention to the Volterra equation (\ref{f7}). For simplicity of notation we
shall drop in what follows the suffix max appearing in (\ref{f7}). After
successive iterations, we encounter the presence of infrared divergent terms in
the perturbative series of $\Pi$:

\bea
\label{ppd}
\Pi(p)=-p^2+\frac\alpha2T^2&-&\alpha^2 T^2\left(\frac{T}{p}\right)+\alpha^3
T^2\left(\frac{T}{p}\right)^2ln\left(\frac{p}{\lambda}\right)
-\frac76\alpha^4
T^2\left(\frac{T}{p}\right)^3\frac{p}{\lambda}+\nonumber\\
&&+\frac{19}{36}\alpha^5 T^2\left(\frac{T}{p}\right)^4
\left(\frac{p}{\lambda}\right)^2-
\frac{149}{540}\alpha^6T^2\left(\frac Tp\right)^5\left(\frac p\lambda\right)^3+
\dots
\eea
where $\alpha\equiv\frac{g^2}{12\pi^3}$.

With the help of this result and using equation (\ref{tl}), we find that the
perturbative expansion of the pressure exhibits power infrared divergences
given by:

\beq
\label{prd}
P_{\rm div}=-\frac{T^6}{96\pi^3}\left[\alpha^3
ln\frac{T}{\lambda}-\frac76\alpha^4\left(\frac{T}{\lambda}\right)
+\frac{19}{36}\alpha^5\left(\frac{T}{\lambda}\right)^2
-\frac{149}{540}\alpha^6\left(\frac{T}{\lambda}\right)^3
+\dots\right]
\eeq

Despite the fact that equation (\ref{f7}) cannot be solved perturbatively
because the iterations yield infrared divergent terms, we will show that
nevertheless, a non-perturbative solution does exist. To this end, it is
convenient to define:

\beq
\label{fx}
x\equiv\frac{\alpha T}{p}\qquad;\qquad f(x)\equiv\frac\Pi{p^2}
\eeq

Then (\ref{f7}) reduces, dropping the inhomogeneous $\lambda^3$ term,
to the following integral equation:

\beq
\label{f11}
f(x)=-1+\frac x3-\frac{x^2}{2\alpha}
-x^2\int_\alpha^x\frac{dy}{y^2}f(y)
-x^4\int_x^{\frac{\alpha T}{\lambda}}\frac{dy}{y^4}f(y)
\eeq

Perturbation theory now corresponds to solving (\ref{f11}) by a power series in
$x$, a procedure which, as we have seen, yields power infrared divergences.

But the differential equation which follows from (\ref{f11}):

\beq
x^2f''-5xf'+(8-2x)f=x-8
\eeq
has a well behaved solution. The complementary one involves modified Bessel
functions\cite{ref10} and two constants:

\beq
f_c(x)=Ax^3I_2(\sqrt{8x})+Bx^3K_2(\sqrt{8x})
\eeq

Using standard methods we can then find a particular solution in terms of
these modified Bessel functions. The general solution, consisting of the sum
of $f_c(x)$ and the particular one has the form:

\bea
f(x)&=&\left[A+2\int_x^{\frac{\alpha T}\lambda}dt
\left(\frac8{t^4}-\frac1{t^3}\right)
K_2(\sqrt{8t})
\right]x^3I_2(\sqrt{8x})+\nonumber\\
&&+\left[B+2\int^x_\alpha dt
\left(\frac8{t^4}-\frac1{t^3}\right)
I_2(\sqrt{8t})
\right]x^3K_2(\sqrt{8x})
\eea

Substituting this into the integral equation (\ref{f11}), fixes the constants
$A$
and $B$ in terms of the parameters $\frac{\alpha T}\lambda$ and $\alpha$. Since
$I_2$ grows exponentially at large $x$, it would produce a divergence in the
integral equation as $\lambda\rightarrow0$, unless $A=0$ in this limit. Indeed,
the consistency conditions on $f(x)$ at $x=\frac{\alpha T}\lambda$ and
$x=\alpha$, demand $A$ to vanish when $\lambda\rightarrow0$,
while $B$ becomes:

\beq
B=\frac1{K_3(\sqrt{8\alpha})}\left[2I_3(\sqrt{8\alpha})
\int_\alpha^\infty dt
\left(\frac8{t^4}-\frac1{t^3}\right)
K_2(\sqrt{8t})-\sqrt{\frac8{\alpha^7}}\,\right]
\eeq

Then, since $K_2$ decreases exponentially at large $x$, it is not difficult to
show that $f(x)$ is a decreasing function of $x$ in this domain. Because large
values of $x$ correspond to small values of the momenta [see eq.(\ref{fx})],
this behavior
leads to a convergent integral in (\ref{tl}) as $\lambda\rightarrow0$. Thus,
the pressure calculated in the ladder approximation remains finite in the
infrared domain. This also happens in the case of the Yang-Mills theory to
which we now turn.

\section{The Yang-Mills Theory}
\label{ym}

We now consider the leading infrared behavior of the pressure in
thermal Yang-Mills theory. Since the $\tilde\Pi_{00}(p\rightarrow0,p_0=0)$
component of
the polarization tensor is non-zero, the longitudinal gluons are screened at
large distances. As a result, the infrared problem is related in this case
only to the behavior of the transverse part of the gluon propagator. Using
the fact that at finite temperatures, the leading infrared contributions arise
from terms with zero frequencies, we can reduce our problem to a study in
$3$-dimensional Euclidian Yang-Mills theory.

Although the analysis is now more complicated due to the presence of the
$4$-gluon couplings and ghosts, it proceeds in parallel with that
described previously in the Section \ref{scalar}.
In particular, in order to derive the
expression for the pressure in the ladder approximation, we start from a basic
relation which is the analogue of eq. (\ref{f1}). By a reasoning similar to
that used in deriving eq. (\ref{tl}), we find that the expression giving the
leading infrared contributions to the pressure has the form:

\beq
\label{fp}
P=P_0-T\int_\lambda^T\frac{d^3p}{(2\pi)^3}\frac1{p^2}\left\{
\frac18\tilde\Pi^{aa}_{ii}(p)-\frac14\tilde\Sigma^{aa}(p)+
\frac1{16}\frac{T^2}{2\pi^2p^2}V^{abcc}_{ijkk}\tilde\Pi^{ab}_{ij}(p)+
\dots\right\}
\eeq

This relation is represented graphically in Fig. (3). Here $\tilde\Pi$ and
$\tilde\Sigma$ denote respectively the gluon and ghost self-energy functions,
which must be calculated consistently in the ladder approximation. The vertex
$V$ stands for the bare gluon four-point function and dots indicate infrared
convergent contributions associated with two-loop diagrams. Note that
expression (\ref{fp}) is much simpler than the exact formula
giving the pressure in the Yang-Mills theory\cite{kal}.

We now proceed to investigate, in the Feynman gauge, the infrared behavior of
the gluon polarization tensor and the ghost self-energy function.
The corresponding Schwinger-Dyson (S-D)
equations in the ladder approximation are shown in Fig. (4).

It is well known that any approximation of the S-D equations may impose severe
constraints in the case of a gauge theory. The reason is that the Ward
identities, which reflect its underlying gauge invariance, are satisfied only
when we take into account all the relevant contributions, order by order in
perturbation theory. In particular, the transversality property of the
polarization tensor is guaranteed only by the full set of S-D equations. Since
it is impossible to implement this program in practice, we follow
the procedure adopted by Mandelstam\cite{ref5}, neglecting the longitudinal
terms which might arise in connection
with the approximate set of S-D equations.
This is obviously correct to lowest order in perturbation theory, where our
polarization tensor is manifestly transverse. In higher orders, some of the
contributions associated with the exact $3$-point and $4$-point gluon vertices
will cancel these longitudinal terms. Since in the ladder approximation the
vertex corrections are also neglected, the above procedure is justified and, as
we shall see, a consistent solution of the approximate set of S-D equations
does exist.

Thus, using the transversality property of the gluon polarization tensor
\hbox{$\tilde\Pi^{ab}_{ij}=\delta^{ab}\tilde\Pi_{ij}$},we can write it in the
form:

\beq
\tilde\Pi_{ij}=\left(\delta_{ij}-\frac{p_i p_j}{p^2}\right)(\Pi_T+p^2)
\eeq

In this way, we arrive at the following set of coupled integral equations which
relate the gluon polarization tensor to the ghost self-energy
$\tilde\Sigma^{ab}=\delta^{ab}(\Sigma+p^2)$:

\bea
\label{f17}
\Pi_T&=&-p^2-\frac{11g^2NT}{48\pi^3}\int\frac{d^3 k}{k^2}
-\frac{g^2NT}{16\pi^3}\int
\frac{d^3 k}{k^4}\frac{2(2k^2+3p^2)({\bf p}\cdot {\bf k})^2-
p^2k^2(6p^2+7k^2)}{p^2({\bf p}+{\bf k})^2}-\nonumber\\
&&-\frac{g^2NT}{24\pi^3}\int\frac{d^3k}{k^4}\Pi_T(k)
+\frac{g^2NT}{8\pi^3}
\int\frac{d^3 k}{k^6}\frac{5p^2k^2-3({\bf p}\cdot
{\bf k})^2}{p^2({\bf p}+{\bf k})^2}(p^2+k^2)\Pi_T(k)+\nonumber\\
&&+\frac{g^2NT}{8\pi^3}\int\frac{d^3 k}{k^4}
\frac{p^2k^2-({\bf p}\cdot {\bf k})^2}{p^2({\bf p}+{\bf k})^2}\Sigma(k)\\
\label{f18}
\nonumber\\
\Sigma(p)&=&-p^2-\frac{g^2NT}{8\pi^3}\int\frac{d^3 k}{k^4}
\frac{({\bf k}\cdot {\bf p})^2}{({\bf p}+{\bf k})^2}+
\frac{g^2NT}{8\pi^3}\int\frac{d^3 k}{k^4}
\frac{({\bf p}\cdot {\bf k})}{({\bf p}+{\bf k})^2}\Sigma(k)-\nonumber\\
&&-\frac{g^2NT}{8\pi^3}\int\frac{d^3 k}{k^6}\frac{p^2k^2-({\bf p}\cdot
{\bf k})^2}{({\bf p}+{\bf k})^2}\Pi_T(k)
\eea

The first iteration of this system yields, as required, the lowest order
perturbative contributions to the gluon and ghost self-energy functions. In
higher orders the
perturbative series of these self-energy functions lead, via equation
(\ref{fp}), to the presence of power infrared divergences in the perturbative
expansion of the pressure. These features are rather similar to the ones
exhibited in Section \ref{scalar} by the thermal scalar field [see eq.
(\ref{ppd}) and (\ref{prd})].

Analogously to the scalar case, we will now show that a well behaved
non-perturbative solution for the above set of integral equations does exist.
In order to be able to solve it analytically, we will use consistently in the
numerators and denominators of equations (\ref{f17}) and (\ref{f18}), the
maximization procedure described in the previous case. As we have seen, in the
high temperature domain this
procedure simplifies the equations, without modifying the qualitative features
of their solutions. Then, it is convenient to define the dimensionless
quantities:

\bea
\label{f19}
\alpha\equiv\frac{g^2N}{3\pi^2}\qquad&;&\qquad x\equiv\frac{\alpha T}{p}\\
\label{ff}
F(x)\equiv\frac{\Pi_T}{p^2}\qquad&;&\qquad G(x)\equiv\frac{\Sigma}{p^2}
\eea

In this way, after performing the angular integrations, we find the following
set of Volterra integral equations:

\bea
\label{f20}
F(x)&=&-1-\frac{13}{12}x-Hx^2+x^2\int_\alpha^x\frac{dy}{y^2}G(y)
+x^4\int_x^\frac{\alpha T}{\lambda}\frac{dy}{y^4}G(y)\\
\label{f21}
G(x)&=&-\left(1+\frac{\alpha}2\right)+x-\int_\alpha^x dyF(y)
-x^2\int_x^\frac{\alpha T}{\lambda}\frac{dy}{y^2}F(y)
\eea
where

\beq
\label{eqh}
H=\frac1{2\alpha}+
\frac{11}2\int_\alpha^{\frac{\alpha T}\lambda}\frac{dy}{y^2}F(y)
\eeq

Perturbation theory corresponds to solving (\ref{f20}) and (\ref{f21}) by a
power series in x. On the other hand, the relevant momenta in the infrared
domain are such that $p<\alpha T$, which correspond to large values of x.

We now consider the differential equations which follow from (\ref{f20}) and
(\ref{f21}):

\bea
\label{f22}
x^2F''(x)-5xF'(x)+8F(x)+2xG(x)&=&-8-\frac{39}{12}x\\
\label{f23}
xG''(x)-G'(x)-2F(x)&=&-1
\eea

To obtain the solution of this set of coupled equations, we use (\ref{f22}) to
express $G(x)$ in function of $F(x)$ and its derivatives. Substituting this
into
(\ref{f23}), we get the following fourth order differential equation:

\beq
\label{f24}
F''''(x)-\frac{4}{x}F'''(x)+\frac{12}{x^2}F''(x)-\frac{24}{x^3}F'(x)
+\left(\frac{24}{x^4}+\frac{4}{x^2}\right)F(x)=\frac{2}{x^2}-\frac{24}{x^4}
\eeq
which describes the infrared behavior of our system.

The complementary solution of the homogeneous part of equation (\ref{f24}) can
be expressed in terms of the modified Bessel functions with complex argument.
Using the functional relations satisfied by these functions\cite{ref10} it is
straightforward to show that:

\beq
\label{f25}
F_c(x)=ix^2[C_1I_2(\sqrt{8ix})+C_2K_2(\sqrt{8ix})]+ {\rm complex\ conjugate}
\eeq
where $C_1$ and $C_2$ are complex constants to be determined.

In order to obtain a particular solution of (\ref{f24}), we employ the method
of variation of the parameters, which is described in Appendix \ref{aa}.
The general
solution, consisting of the sum of $F_c(x)$ and $\tilde F(x)$ [equation
(\ref{fa7})] is given by:

\bea
\label{f26}
F(x)&=&ix^2\left\{\left[C_1+\int_x^\frac{\alpha
T}{\lambda}dt\left(\frac{1}{t^2}-\frac{12}{t^4}\right)
K_2(\sqrt{8it})\right]I_2(\sqrt{8ix})\right.\nonumber\\
&&\left.+\left[C_2+\int_\alpha^x dt\left(\frac{1}{t^2}
-\frac{12}{t^4}\right)I_2(\sqrt{8it})\right]
K_2(\sqrt{8ix})\right\}+ {\rm c.c.}
\eea

Since $I_2$ grows exponentially for large values of its argument, this result
would produce a divergence as $\lambda\rightarrow 0$ in the integral
equations (\ref{f20}) and (\ref{f21}), unless $C_1$ vanishes. In Appendix
\ref{bb}, we
discuss the consistency conditions which follow from these equations in the
limit $\lambda\rightarrow 0$. We find that, indeed, these conditions
require the vanishing of the constant $C_1$. Futhermore, they fix the constant
$C_2$ in terms of $\alpha$, but the explicit expression has a rather
complicated form involving the modified Bessel functions [see eq. (\ref{eqd})].

{}From the behavior of $F(x)$ and $G(x)$ for large values of $x$, given by
(\ref{fb4}) and (\ref{fb5}), and using eqs. (\ref{f19},\ref{ff}), we see that
$\Pi_T(p)$
and $\Sigma(p)$ will be proportional to $p^2$ for small values of the momenta:

\beq
\Pi_T(p)\cong\frac12p^2\qquad;\qquad\Sigma(p)\cong-\frac{39}{24}p^2\qquad\qquad
[p<<\alpha T]
\eeq

This behavior will ensure the convergence of the integral in
eq. (\ref{fp}) as $\lambda\rightarrow0$.
Hence we find that the leading contributions to the pressure, when calculated
in the ladder approximation of the $SU(N)$ Yang-Mills theory, are finite in the
infrared domain.

In conclusion, we stress that our analysis was restricted to the ladder
approximation, in which case the
set of Schwinger-Dyson equations remains linear, allowing for an
analytic solution. However, there are many other important contributions which
must be taken into account when studying the infrared behavior of the
pressure. These include the set of diagrams with crossed ghost and gluon
lines, which can be reduced in the large $N$-limit to the class of
planar diagrams . Furthermore, one must also consider additional self-energy
and vertex corrections, which are relevant for the appearance of the effective
coupling constant $g(T)$. It is conceivable that the inclusion of these
contributions, may indicate the presence of new collective phenomena
associated with the infrared behavior of the pressure. Nevertheless, in view
of the above results, we believe that the
cancellation of the infrared divergences
is a possibility which deserves further investigation.

\acknowledgements

We would like to thank CNPq and FAPESP (Brasil) for support and to A. F.
Granero for his assistance. J. F. is grateful to Prof. J. C. Taylor for a
valuable correspondence and for reading the manuscript.

\newpage
\appendix{}
\label{aa}

Here we derive a particular solution of eq. (\ref{f24}), using the method of
variation of the parameters. The complementary solution was given in eq.
(\ref{f25}), so we assume a particular one of the form:

\beq
\label{fa1}
\tilde F(x)=Z_1(x)x^2I_2(\sqrt{8ix})+Z_2(x)x^2K_2(\sqrt{8ix})+{\rm c.c.}
\eeq
where $Z_1(x)$ and $Z_2(x)$ are two complex functions to be determined
consistently. We have to our disposal four real functions, but only one
condition, so that we are free to impose three more conditions. We choose
these additional conditions as follows:

\bea
\label{fa2}
Z_1'(x)x^2I_2(\sqrt{8ix})+Z_2'(x)x^2K_2(\sqrt{8ix})+{\rm c.c.}=0\\
\label{fa3}
Z_1'(x)\left[x^2I_2(\sqrt{8ix})\right]'+
Z_2'(x)\left[x^2K_2(\sqrt{8ix})\right]'+{\rm c.c.}=0\\
\label{fa4}
Z_1'(x)\left[x^2I_2(\sqrt{8ix})\right]''+
Z_2'(x)\left[x^2K_2(\sqrt{8ix})\right]''+{\rm c.c.}=0
\eea

Then, the expressions giving $\tilde F'(x)$, $\tilde F''(x)$, $\tilde F'''(x)$
and $\tilde F''''(x)$ simplify considerably. We now impose the basic condition
that (\ref{fa1}) be a solution of (\ref{f24}). Thus we substitute these
expressions into (\ref{f24}) and obtain an identity. Since
$x^2I_2(\sqrt{8ix})$, $x^2K_2(\sqrt{8ix})$ and their complex conjugates are
solutions of the homogeneous equation, this identity reduces to:

\beq
Z_1'(x)\left[x^2I_2(\sqrt{8ix})\right]'''+
Z_2'(x)\left[x^2K_2(\sqrt{8ix})\right]'''+{\rm c.c.}=\frac{2x^2-24}{x^4}
\eeq

Together with the auxiliary conditions (\ref{fa2}), (\ref{fa3}) and
(\ref{fa4}), we have a system of equations which determinate the complex
functions
$Z_1'(x)$ and $Z_2'(x)$. To this end we use the functional relations
satisfied by $K_2(z)$ and $I_2(z)$, together with the fact that their Wronskian
is $\frac1z$. Then, after a straightforward calculation we obtain:

\bea
Z_1'(x)&=&-i\left(\frac1{x^2}-\frac{12}{x^4}\right)K_2(\sqrt{8ix})\nonumber\\
Z_2'(x)&=&i\left(\frac1{x^2}-\frac{12}{x^4}\right)I_2(\sqrt{8ix})
\eea

Integrating these relations and substituting the result into (\ref{fa1}), we
find that the particular solution $\tilde F(x)$ can be expressed in the form:

\bea
\label{fa7}
\tilde F(x)&=&ix^2\left\{I_2(\sqrt{8ix})\int_x^{\frac{\alpha T}\lambda}dt
\left(\frac1{t^2}-\frac{12}{t^4}\right)K_2(\sqrt{8it})+\right.\nonumber\\
&&\left.+K_2(\sqrt{8ix})\int_\alpha^xdt
\left(\frac1{t^2}-\frac{12}{t^4}\right)I_2(\sqrt{8it})\right\}+{\rm c.c.}
\eea

With the help of this relation, the expression (\ref{f26}) for the general
solution $F(x)$ can now be easily deduced. Using the series representation of
the modified Bessel functions\cite{ref10}, it can be verified $F(x)$ satisfies
the perturbative boundary condition $F(0)=-1$.

\appendix{}
\label{bb}

We discuss here the determination of the constants $C_1,C_2$ in the infrared
limit $\lambda\rightarrow 0$, and obtain the asymptotic forms of $F(x)$ and
$G(x)$ for large values of $x$.

To this end we substitute (\ref{f22}) and (\ref{f23}) back into the original
integral equations (\ref{f20}) and (\ref{f21}). When the expression (\ref{f26})
for $F(x)$ is used in these equations, we obtain two relations which must be
satisfied identically for all values of $x$. The constants $C_1$ and $C_2$ are
fixed by the consistency conditions which arise from these relations at the
points $x=\frac{\alpha T}{\lambda}$ and $x=\alpha$, respectively.
Using the properties of the
modified Bessel functions\cite{ref10}, it is straightforward to show in the
limit $\lambda\rightarrow 0$, that these conditions demand $C_1$ to vanish.
Then $C_2$ is determined by the system:

\bea
\left.\frac{x^3}{2}\frac{d}{dx}
\left(\frac{G(x)}{x^2}\right)\right|_{x=\alpha}&=&1\nonumber\\
\left.\frac{x^3}{2}\frac{d}{dx}\left(\frac{F(x)}{x^4}\right)\right|_{x=\alpha}
&=&H+\frac{13}{8\alpha}+\frac{2}{\alpha^2}
\eea

This set of equations fixes the complex constant $C_2$ uniquely, since the
discriminant $D$ of the system formed by $C_2$ and $C_2^*$ is non-vanishing. To
evaluate it, we use the expression (\ref{f26}) which gives the general
solution $F(x)$ in terms of the modified Bessel functions, together with the
fact that $C_1=0$. Futhermore, we express $G(x)$ [eq.(\ref{f22})] in terms of
$F(x)$ and its derivatives, as well as $H$ [eq. (\ref{eqh})] in terms of an
integral over $F(x)$. These factors make the explicit expression of the
discriminant have a rather complicated form involving the modified Bessel
functions. Using the relations satisfied by these functions, we find after a
straightforward calculation that:

\bea
\label{eqd}
D&=&-i\alpha\left(2\alpha+\frac{11}4\right)K_2(\beta)K_2(\beta^*)-
i\alpha^2\left(\alpha-11\right)K_2'(\beta)K_2'(\beta^*)+\nonumber\\
&&+i\left[\frac\alpha\beta\left(\frac{33}4-11i\alpha+\frac32\alpha+4i\alpha^2
-11\alpha^2\right)
K_2'(\beta)K_2(\beta^*)+{\rm c.c.}\right]
\eea
where $\beta\equiv\sqrt{8i\alpha}$ and $K_2'(z)\equiv\frac d{dz}K_2(z)$.

Since $C_1$ vanishes, we see from equation (\ref{f25}) that the complementary
solution decreases exponentially for large values of $x$. Then, the asymptotic
form of $F(x)$ is determined by $\tilde F(x)$, which in the infrared limit
$\lambda\rightarrow 0$, becomes:

\beq
\label{b2}
F(x)\cong ix^2\left\{K_2(\sqrt{8ix})\int_\alpha^x\frac{dt}{t^2}I_2(\sqrt{8it})
+I_2(\sqrt{8ix})\int_x^\infty\frac{dt}{t^2}K_2(\sqrt{8it})\right\}+{\rm c.c.}
\eeq

In order to evaluate $F(x)$ for large values of $x$, we use the asymptotic
expansions of the modified Bessel functions\cite{ref10}:

\bea
I_2(z)&\cong&\frac{1}{\sqrt{2\pi z}}e^z\nonumber\\
K_2(z)&\cong&\sqrt{\frac{\pi}{2z}}e^{-z}
\qquad
[|z|>>1]
\eea

Then, the leading contributions arising from the integrations in (\ref{b2})
can be calculated explicitly. In this way, we find that the asymptotic
behavior of $F(x)$, for large values of $x$, is given by:

\bea
\label{fb4}
F(x)\cong\frac{1}{2}\qquad[x>>1]
\eea

With the help of this result, we obtain from (\ref{f22}) the asymptotic form
of $G(x)$ for large values of $x$:

\beq
\label{fb5}
G(x)\cong-\frac{39}{24}\qquad[x>>1]
\eeq

Since $x=\frac{\alpha T}{p}$, these expressions describe equivalently the
behavior of our solutions for small values of the momenta.

\appendix{}
\label{cc}

Here we present a very simple model which illustrates a possible mechanism for
the cancellation of the infrared divergences, outside the ladder approximation.
The model has some similarities
with the magnetic sector of the thermal Yang-Mills theory. We consider the
following expression for the pressure:

\beq
P=P_0+cT\sum_{p_0}\int\frac{d^3p}{p^2+p_0^2}\tilde\Pi(p,p_0)
\eeq

The perturbative series of $\tilde\Pi(p\rightarrow0,p_0=0)$ is given by:

\beq
\label{c2}
\tilde\Pi(p)=-ag^2pT+b\left(g^2T\right)^2+\dots
\eeq
where $a$, $b$ and $c$ are positive constants. We see that a thermal mass of
order $g^2T$ is generated, a possibility which might occur also in the case of
the magnetic mass. Our model assumes, for definiteness, that the contributions
to $\tilde\Pi(p)$ exponentiate. Then, (\ref{c2}) can be written as follows:

\beq
\tilde\Pi(p)=\frac{a^2}{2b}p^2\left(e^{-\frac{\alpha T}p}-1\right)
\eeq
where $\alpha\equiv\frac{2b}ag^2$.

Proceeding in parallel with previous cases, we obtain that the expression
describing the dominant infrared behavior of the pressure has the form:

\beq
\label{c4}
P\cong2\pi\frac{a^2c}bT\int_\lambda^Tp^2e^{-\frac{\alpha T}p}dp
\eeq

Using this form, we find that the perturbative expansion of the pressure
contains power infrared divergences given by:

\beq
P_{\rm div}\cong2\pi\frac{a^2c}bT^4\left[-\frac{\alpha^3}{3!}ln\left(\frac
T\lambda\right)+\frac{\alpha^4}{4!}\frac T\lambda-\frac{\alpha^5}{5!2}
\left(\frac T\lambda\right)^2+\dots\right]
\eeq

Note that this behavior is similar to the one shown by the perturbative thermal
Yang-Mills theory.

However, the full expression (\ref{c4}) is well behaved in the infrared limit
$\lambda\rightarrow0$. Indeed, introducing the variable $x=\frac{\alpha T}p$,
we find in this limit that $P$ becomes:

\beq
P\cong2\pi\frac{a^2c}bT^4\alpha^3\int_\alpha^\infty\frac{dx}{x^4}e^{-x}
\eeq

This can be evaluated in terms of the exponential-integral
function\cite{ref10}:

\beq
P\cong\frac\pi3\frac{a^2c}bT^4\left[\alpha^3{\rm Ei}(-\alpha)
+(2-\alpha+\alpha^2)e^{-\alpha}\right]
\eeq

Therefore, in our model the complete expression giving the pressure in the
infrared domain in a well behaved function of $\alpha$.

\figure{The Schwinger-Dyson equation for the scalar self-energy
function in the ladder approximation.}

\figure{(a) A general contribution of the thermal scalar field to the pressure
in the ladder approximation and (b) its graphical representation as given by
equation (\ref{f2}).}

\figure{Graphical representation of equation (\ref{fp}). Wavy lines denote bare
gluons and dashed lines represent bare ghost particles. Combinatorial factors
are shown explicitly in the diagrams.}

\figure{(a) The Schwinger-Dyson equation for the gluon polarization tensor in
the ladder approximation and (b) the corresponding equation for the ghost
two-point function.}


\begin{references}

\bibitem{ref1}
J. Kapusta, Finite Temperature Field Theory (Cambridge University Press, 1989)
and references cited therein.

\bibitem{ref2}
A. D. Linde, \pl{B96}{289}{80}

\bibitem{ref3}
O. K. Kalashnikov and V. V. Klimov,
\fp{31}{613}{83}

\bibitem{ref4}
R. Jackiw and S. Templeton, \pr{D82}{615}{81}

\bibitem{ref5}
S. Mandelstam, \pr{D20}{3223}{79}

\bibitem{ref6}
K. Kajantie and J. Kapusta, \ap{160}{477}{85}

\bibitem{ref8}
E. Braaten and R. D. Pisarski, \np{B337}{569}{90}

\bibitem{ref9}
J. Frenkel, A. V. Saa and J. C. Taylor, The Pressure in Thermal Scalar Field
Theory to Three Loop Order, to be published in {\it Phys. Rev.\/} D.

\bibitem{ref10}
I. S. Gradshteyn and I. M. Ryzhik, Tables of Integrals Series
and Products (Academic Press, New York, 1980).

\bibitem{kal}
O. K. Kalashnikov, \fp{32}{525}{84}

\end{references}
\end{document}